\documentclass[12pt]{iopart}

\usepackage{graphicx}
\begin{document}

\title{The conjunction fallacy and interference effects}

\author{Riccardo Franco
\footnote[3]{To whom correspondence should be addressed
riccardo.franco@csi.it}}
\address{Dipartimento di Fisica and U.d.R. I.N.F.M., Politecnico di Torino
C.so Duca degli Abruzzi 24, I-10129 Torino, Italia}

\date{\today}

\begin{abstract}
In the present article we consider the conjunction fallacy, a well known cognitive heuristic experimentally tested in
cognitive science, which occurs for intuitive judgments in situations of bounded rationality. We show that the quantum
formalism can be used to describe in a very simple way this fallacy in terms of interference effect. We evidence that
the quantum formalism leads quite naturally to violations of Bayes' rule when considering the probability of the
conjunction of two events. Thus we suggest that in cognitive science the formalism of quantum mechanics can be used to
describe a \textit{quantum regime}, the bounded-rationality regime, where the cognitive heuristics are valid.
\end{abstract}

\pacs{
} \maketitle

%
\section{Introduction}
This article addresses two main directions of research: the investigation of how the quantum formalism is compatible
with Bayes' rule of classic probability theory, and the attempt to describe with the quantum formalism systems and
situations very different from the microscopic particles. A number of attempts has been done to apply the formalism of
quantum mechanics to research fields different from quantum physics, for example in the study of rational ignorance
\cite{Rfranco_rat_ign_1} and of semantical analysis \cite{contexts_q}. Quantum mechanics, for its counterintuitive
predictions, seems to provide a good formalism to describe puzzling effects of contextuality. In the present article,
we will try to describe within the quantum formalism an important heuristic of cognitive science, the conjunction
fallacy \cite{Kahn-Tversky-conj-fall}. This heuristic is valid in regime of bounded rationality, which is characterized
by cognitive limitations of both knowledge and cognitive capacity. Bounded rationality \cite{bounded} is a central
theme in behavioral economics and it concerns with the ways in which the actual decision-making process influences
agents' decisions. A first attempt to describe this heuristic in terms of quantum formalism has been done in
\cite{Rfranco_conj1}, without evidencing the importance of interference effects.

This article is organized in order to be readable both from quantum physicists and from experts of cognitive science.
In section \ref{section:formalism and test} we introduce the basic notation of quantum mechanics, and we show in
\ref{section:repeated_test} that the quantum formalism describing two non-commuting observables leads to violations of
Bayes' rule. In section \ref{section:bounded rationality} we describe the answers to a question in bounded-rationality
regime in terms of vector state and density matrix of the quantum formalism. Finally, in section
\ref{section:conjunction} we show how the quantum formalism can naturally describe the conjunction fallacy.
%

The main results of this article are: 1) tests on non-commuting observables lead to violations of Bayes' rule; 2) the
opinion-state of an agent for simple questions with only two possible answers can be represented, in
bounded-rationality regime, by a qubit state; 3) the different questions in bounded-rationality regime can be formally
written as operators acting on the qubit states; 4) the explicit answer of an agent to a question in regime of bounded
rationality can be described as a collapse of the opinion state onto an eigenvector of the corresponding operator; 5)
The probability relevant to a question $A$, when analyzed in terms of the probability relevant to a second question $B$
(corresponding to a non-commuting operator) evidences the violation of Bayes' rule. The conjunction fallacy thus
results as a consequence of this general fact.

In conclusion, we present a very general  and abstract formalism
which seems to describe the heuristic of conjunction fallacy. We
think that a similar study can be done for other heuristics of
cognitive science (this will be presented in new papers). Thus
these heuristics could be simple applications of a general theory
describing the bounded-rationality regime, which probably will
lead to new interesting predictions. This could confirm the
hypothesis  that the processes of intuitive judgement could
involve mechanisms at a quantum level in the brain.
\section{Quantum basic formalism}\label{section:formalism and test}
%
We first introduce the standard bra-ket notation usually used in
quantum mechanics,  introduced by Dirac \cite{Dirac}, and then the
density matrix formalism. In particular, we focus our attention on
the concept of qubit. In the simplest situation, a quantum state
is defined by a ket $|s\rangle$, which is a vector in a complex
separable Hilbert space $H$. If the dimension of $H$ is 2, the
state describes a qubit, which is the unit of quantum information.
Any quantum system prepared identically to $|s\rangle$ is
described by the same ket $|s\rangle$.

In quantum mechanics, we call a measurable  quantity an
\textit{observable}, mathematically described by an operator, for
example $\widehat{A}$, with the important requirement that it is
hermitian: $\widehat{A} = \widehat{A}^{\dag}$, where
$\widehat{A}^{\dag}$ is the conjugate transpose. In the case of a
single qubit,  any observable $\widehat{A}$ has two real
eigenvalues $a_0$ and $a_1$ and two corresponding eigenvectors
$|a_0\rangle$ and $|a_1\rangle$. Another property of hermitian
operators is that its eigenvectors, if normalized, form an
orthonormal basis, that is
\begin{equation}\label{orthonormality}
\langle a_i|a_j\rangle = \delta(i,j)\,, \end{equation}
where $i,j=0,1$ and $\delta(i,j)$ is the Kroneker delta, equal to
1 if $i=j$ and null otherwise.
Given such a basis in the Hilbert space, we can write them in
components as
\begin{equation}\label{vector}
|a_0\rangle = \left (
  \begin{tabular}{c}
    1 \\
    0
    \end{tabular} \right ) \,,\,\,
|a_1\rangle = \left (
  \begin{tabular}{c}
    0 \\
    1
    \end{tabular} \right )\,\,,
\end{equation}
representing the quantum analogue to the two possible values 0 and
1 of a classical bit.
%
An important difference is that in the quantum case a state can be
in a linear superposition of 0 and 1, that is
\begin{equation}\label{superposition_A} |s\rangle=s_0 |a_0\rangle + s_1 |a_1\rangle\,,
\end{equation}
with $s_0$ and $s_1$ complex numbers. We also say that the  state
$|s\rangle$ is a superposition of the eigenstates $|a_i\rangle$.
In the vector representation generated by formula (\ref{vector}),
the ket $|s\rangle$ and its dual vector, the bra $\langle s|$, can
be written respectively as
\begin{equation}\label{ket_bra_superposition}
|s\rangle= \left (
  \begin{tabular}{c}
    $s_0$ \\
    $s_1$
    \end{tabular} \right )
\,,\,\,\, \langle s|= \left (
  \begin{tabular}{cc}
    $s_0^*$ & $s_1^*$
    \end{tabular} \right )\,.
\end{equation}
Another mathematical object, which is important in order to
describe probabilities, is the inner product, also called
\textit{braket}. In general, the inner product of two kets
$|s\rangle$ and $|s'\rangle$ can be written, in the basis of the
obsevable $\widehat{A}$, as $\langle
 s|s'\rangle=s_0 {s'_0}^*+s_1 {s'_1}^*$, where $s'_i$ are the
components of $|s'\rangle$ in the same basis. Thus the  inner
product of $|s\rangle$ and its dual vector is $\langle
s|s\rangle=|s_0|^2+|s_1|^2$, and it is equal to 1 if the vector is
normalized.
Finally, $|s_i|^2=|\langle a_i|s\rangle|^2$ is the probability
$P(a_i)$ that the measure on the observable $\widehat{A}$ has the
outcome $i=0,1$.
The state $|s\rangle$ is called a \textit{pure} state, and
describes a quantum state for which the preparation is complete: a
preparation is complete when all the compatible observables have
been defined (we will give further a precise definition of
compatible observables).

%
In the most general case, a quantum state is described by the
density matrix $\widehat{\rho}$, which is an hermitian operator
acting on $H$.
The density matrix $\widehat{\rho}$ describes in general  a
\textit{mixed state}, that is a state for which the preparation is
not completely determined. For example, the state may be in the
preparation $|s_1\rangle$ with a probability $P_1$, and in the
preparation $|s_2\rangle$ with a probability $P_2$ (the two
vectors may  be not orthogonal). We also say that the the mixed
state is a (statistical) mixture of the two states (or of the two
preparations):
\begin{equation}\label{mixed}
\widehat{\rho}=\sum_{i}P_i |s_i\rangle \langle s_i|\,.
\end{equation}
In the particular case where there is only one $P_i=1$, we have
$\widehat{\rho}=|s\rangle \langle s|$, that is a pure state. The
opposite situation is when the eigenvalues  of the density matrix
are all equal. In the single-qubit example, they are both $1/2$,
and the resulting operator is the identity matrix acting on the
Hilbert space $H$:
\begin{equation}\label{maximally_mixed}
\widehat{\rho}= \frac{1}{2} \widehat{I}= \frac{1}{2}\left[
    \begin{tabular}{cc}
    1 & 0 \\
    0 & 1
    \end{tabular}\right]\,.
\end{equation}
The resulting state is called maximally mixed, and can be
considered as the situation where the actual knowledge of the
state is null.
%
%
The elements of the single-qubit density matrix $\widehat{\rho}$
can be expressed in the basis of $\widehat{A}$, with $i,j=0,1$, as
\begin{equation}\label{mixed_elements}
\rho_{i,j}=\langle a_i|\widehat{\rho}|a_j\rangle \,.
\end{equation}
where the diagonal elements $\rho_{i,i}$ represent the probability
$P(a_i)$ to measure a certain value $a_i$ of the relevant
observable. An equivalent expression of these probabilities can be
written in terms of the trace-matrix operation:
\begin{equation}\label{probabilities}
P(a_i)=Tr(\widehat{\rho} |a_i\rangle \langle a_i|)=\rho_{i,i}
\end{equation}
This formula is the most general expression of the probability to
measure a value $a_i$ of an observable, when operating on quantum
systems identically prepared in the state $\widehat{\rho}$.
%
The formalism of the density matrix helps us to write in the most
general form the mean value of an observable $\widehat{A}$ as
\begin{equation}\label{mean_value}
\langle \widehat{A} \rangle = Tr(\widehat{\rho} \widehat{A})\,,
\end{equation}
which becomes, by using formula (\ref{mixed_elements}) and the
basis vectors of $\widehat{A}$, $\langle \widehat{A}
\rangle=\sum_i a_i P(a_i)$.
\subsection{Collapse of the state vector}
One of the axioms \cite{Dirac} of quantum mechanics states that,
given an initial mixed  state $\widehat{\rho}$ and an observable
$\widehat{A}$ acting on a discrete Hilbert space, we can define
from the eigenvectors $\{|a_i\rangle\}$ of $\widehat{A}$ the
projection operators $\{|a_i\rangle\langle a_i|\}$, where for a
single qubit $i=0,1$. Thus if the measure of the observable
$\widehat{A}$ is the eigenvalue $a_i$, the state updates as
\begin{equation}\label{collapse}
\widehat{\rho}  \rightarrow |a_i\rangle\langle a_i|
\end{equation}
This general formula is valid for an orthonormal basis
$\{|a_i\rangle\}$, and defines the collapse of the initial state
onto the state vector $|a_i\rangle$.
At first the collapse of the state may seem quite obvious. For
example, given the initial probability distribution $P(a_j)$ given
by equation (\ref{probabilities}) corresponding to the initial
state $\widehat{\rho}$, we have from simple calculations that
\begin{equation}\label{collapse_P1}
P(a_j) \rightarrow P(a_j)=\delta(a_i,a_j)\,.
\end{equation}
This means that, after measuring a certain value $a_i$ of
$\widehat{A}$, the probability that the observable actually has
another value $a_j\neq a_i$ is null. This fact is valid also in
classic probability theory. Nevertheless, we will show in the next
subsection that the collapse leads to violation of classic laws of
probability theory when considering more than one observable.

Finally we note that, in the study of rational ignorance presented
in \cite{Rfranco_rat_ign_1}, the collapse admits a very simple
interpretation. Given an initial opinion state, described by
$\widehat{\rho}$ and a question $\widehat{A}$, after a subject has
given an answer $a_i$, the probability that the repetition of the
same question $\widehat{A}$ in the \textit{same conditions} gives
a different answer is null.
\subsection{Non-commuting operators and Bayes' rule}\label{section:repeated_test}
In quantum mechanics the operators associated to the observables
may not commute: for example, given two operators $\widehat{A}$
and $\widehat{B}$, acting on the same Hilbert space $H$, the
ordered product $\widehat{A}\widehat{B}$ can be different from
$\widehat{B}\widehat{A}$: in this case the two operators do not
commute and we write $[\widehat{A},\widehat{B}]\neq 0$, where
$[\widehat{A},\widehat{B}]=\widehat{A}\widehat{B}-\widehat{B}\widehat{A}$
is the commutator of the two observables. The consequences of this
fact are very important, and lead to violation of Bayes' rule.
Let us consider for simplicity a single-quit system and the
eigenvectors $\{|a_i\rangle\}$ and $\{|b_i\rangle\}$ of
$\widehat{A}$ and $\widehat{B}$ respectively (with $i=0,1$). The
probability of measuring the value $a_i$ or $b_i$ for the
observables $\widehat{A}$ or $\widehat{B}$ respectively is given
by equation (\ref{probabilities}), that is:
\begin{equation}\label{prob_AB}
P(a_i)=Tr(\widehat{\rho} |a_i\rangle \langle a_i|); \,\,
P(b_i)=Tr(\widehat{\rho} |b_i\rangle \langle b_i|)\,.
\end{equation}
We now  consider the conditional probability $P(b_j|a_i)$, defined
as the probability to measure the observable $\widehat{B}$ with
value $b_j$, given the occurrence of a measurement of
$\widehat{A}$ with value $a_i$. In quantum mechanics, the
occurrence of a measurement of $\widehat{A}$ with result $a_i$
means that the actual state is $|a_i\rangle$, independently from
the initial state before the measurement. This is a consequence of
the quantum collapse, and leads to many differences form the
classic case. In quantum mechanics thus we have that
\begin{equation}\label{conditional_p}
P(b_j|a_i)=|\langle b_j|a_i\rangle|^2=P(a_i|b_j)\,.
\end{equation}
Let us now consider the Bayes' rule, which defines the joint
probability to measure contemporarily the values $a_i$ and $b_j$
for observables $A$ and $B$ respectively:
\begin{equation}\label{joint_cond}
P(a_i)P(b_j|a_i)=P(b_j)P(a_i|b_j)=P(a_i,b_j)\,.
\end{equation}
This equation is very important in classical probability theory,
since it links the joint probabilities relevant to $\widehat{A}$
and $\widehat{B}$ to the conditional probabilities.
In quantum mechanics one can not measure contemporarily two
commuting operators. From a formal point of view, this
impossibility is evidenced from the fact that, by using the
equations (\ref{prob_AB}) and (\ref{conditional_p}), we have in
general that
\begin{equation}\label{joint_cond1}
P(a_i)P(b_j|a_i)\neq P(b_j)P(a_i|b_j)
\end{equation}
This means that the joint probability $P(a_i,b_j)$ can not be
univocally defined. What we can rigorously define is
\begin{equation}\label{joint_cond2}
P(a_i \rightarrow b_j)=P(a_i)P(b_j|a_i)\,,
\end{equation}
where $P(a_i \rightarrow b_j)$ is the probability to measure $a_i$
for the observable $\widehat{A}$ and then the answer $b_j$ for the
observable $\widehat{B}$ . Form the previous observation, we have
that $P(a_i \rightarrow b_j) \neq P(b_j \rightarrow a_i)$. We call
$P(a_i \rightarrow b_j)$ the \textit{consecutive probability} to
measure $a_i$ and then $b_j$.
Equation (\ref{joint_cond1}) evidences that in quantum mechanics the Bayes' rule is violated. Many of the paradoxical
results of quantum mechanics  are due to this violation. In the present article, we will focus our attention on the
conjunction fallacy, which we will study in the next sections.
%
%
\section{Bounded rationality and Hilbert spaces }\label{section:bounded rationality}
The bounded rationality  \cite{bounded} is a property of an agent
(a person which makes decisions) that behaves in a manner that is
nearly optimal with respect to its goals and resources. In
general, an agent acts in bounded-rationality regime when there is
a limited time in which to make decisions, or when he is also
limited by schemas and other decisional limitations. As a result,
the decisions are not fully thought through and they are rational
only  within limits such as time and cognitive capability. There
are two major causes of bounded rationality, the limitations of
the human mind, and the structure within which the mind operates.
This impacts decision models that assume us to be fully rational:
for example when calculating expected utility, it happens that
people do not make the best choices. Since the effects of bounded
rationality are counterintuitive and may violate the classical
probability theory (and the Bayes' rule), we will often speak of
\textit{bounded-rationality regime} as the set of situations where
the bounded rationality is an actual property.

We will show that some typical behaviors of the bounded-rationality regime can be described in a very effective way by
the quantum formalism. We will study from a statistical point of view the opinion state of agents having the same
initial information. In particular, we will assume that the opinion state of an agent can be represented as a qubit
state, that is in terms of a density matrix $\widehat{\rho}$ or, in simple cases,  of a ket $|s\rangle$ in a Hilbert
space $H$ of dimension 2.
As in quantum mechanics experiments, it  is important to define carefully the preparation of the opinion state. Every
previous information given to an agent before performing a test can be considered as the preparation of the opinion
state. When we repeat a test on more agents, it is important that their opinion state is (at least in  theory)
\textit{identically prepared}. We note here that it is not easy to prepare the opinion state of a number of people in
an identical state. Nevertheless, the quantum formalism can help us with the concept of mixed state.

The basic test in the context of bounded rationality is a
question. We consider a question $A$ for which the possible
answers can only be 0 or 1 (false or true), and we associate it to
an operator $\widehat{A}$ acting on the Hilbert space $H$. Like in
quantum mechanics, the question $A$ is an observable, in the sense
that we can observe an answer: thus, when speaking of questions,
we will consider directly the associated operator $\widehat{A}$.
The answers 0 and 1 are associated to the eigenvalues $a_0=0$ and
$a_1=1$ of $\widehat{A}$, while the eigenvectors $|a_0\rangle$ and
$|a_1\rangle$ correspond to the opinion states relevant to the
answers 0 and 1 respectively:
\begin{equation}\label{eig1}
\widehat{A}|a_0\rangle=0|a_0\rangle=0;\,\,\,\widehat{A}|a_1\rangle=|a_1\rangle\,\,.
\end{equation}
The eigenvectors $|a_0\rangle$ and $|a_1\rangle$ have a very
precise meaning: if the opinion state of an agent can be described
for example by $|a_1\rangle$, this means that the answer to the
question $\widehat{A}$ is 1 with certainty. If we repeat the same
question to many agents in the same opinion state (thus
identically prepared), each agent will give the same answer 1. If
instead the opinion state about the question is definitely 0, then
we have the eigenvector $|a_0\rangle$.
Any observable can be written in the basis of its eigenvectors as
$\widehat{A}=\sum_i a_i |a_i\rangle\langle a_i|$.
A superposition of the opinion states $|a_i\rangle$ about question
$\widehat{A}$ is, like in equation (\ref{superposition_A}),
$|s\rangle=s_0 |a_0\rangle + s_1 |a_1\rangle$, where
$|s_i|^2=|\langle a_i|s\rangle|^2$ is the probability $P(a_i)$
that the agent gives an answer $i=0,1$ to the question
$\widehat{A}$.
In the most general case, the opinion state can be represented as
a density matrix $\widehat{\rho}$. The probability that the answer
to the question $\widehat{A}$ is $a_i$ is given by equation
(\ref{probabilities}): $P(a_i)=Tr(\rho |a_i\rangle \langle a_i|)$.

We note that the formalism introduced is the same used to describe
questions in regime of rational ignorance
\cite{Rfranco_rat_ign_1}, where people choose to remain uninformed
about a question $\widehat{A}$. In fact, the bounded rationality
can be considered as a more general than the rational ignorance,
where the question is preceded by some additional information.
\section{The conjunction fallacy}\label{section:conjunction}
The \textit{conjunction fallacy}  is a well known cognitive heuristic which occurs in bounded rationality when some
specific conditions are assumed to be more probable than the general ones. More precisely, many people tend to ascribe
higher probabilities to the conjunction of two events than to one of the single events. The most often-cited example of
this fallacy originated with Amos Tversky and Daniel Kahneman \cite{Kahn-Tversky-conj-fall} is the case of Linda, which
we will consider carefully in this article. The conjunction fallacy has been later studied in a detailed way
\cite{osherson}, in order to show that the fallacy does not depend by other factors: for example, the interpretation of
expressions like \textit{probability} and \textit{and}.

In general, we consider two dichotomic questions $A$ and $B$, with possible answers $a_0, a_1$ and $b_0, b_1$. The
typical experiments of \cite{Kahn-Tversky-conj-fall} and \cite{osherson} consist in  a preparation of the opinion
state, which provides some information to the agent, and the following question: what is more probable or frequent
between  $a_1$ (or $b_1$) and $a_1$-and-$b_1$. The agents manifest in all these experiments a strict preference for the
answer $a_1$-and-$b_1$: this evidences the conjunction fallacy, since the Bayes' rule (\ref{joint_cond}) entails that
\begin{equation}\label{Conj_fallacy_classic}
P(b_1)=P(a_0)P(b_1|a_0)+P(a_1)P(b_1|a_1)\geq P(a_1)P(b_1|a_1) \,.
\end{equation}
In other words, the conjunction of two events $a_1$ and $b_1$ is always less probable than of one of two events.
Nonetheless, the agents often consider more likely $a_1$-and-$b_1$ than $a_1$ (or $b_1$).
We stress that the experimental results of \cite{Kahn-Tversky-conj-fall} and \cite{osherson} should be considered
carefully: they give a direct information of how many agents consider the difference $P(a_1,b_1)-P(a_1)$ positive, not
of the probabilities relevant to $a_1$-and-$b_1$ and $a_1$ (or $b_1$). The probability $P(a_1,b_1)$ can be obtained
with a different test, where it is asked to the agents if they consider $a_1$-and-$b_1$ true or false, with the
information provided in the preparation of the test.

We now introduce the quantum formalism in order to show that the conjunction fallacy can be described and interpreted
in such a formalism. In particular, we consider two operators $\widehat{A}$ and $\widehat{B}$, associated respectively
to the questions $A$ and $B$. Since $A$ and $B$ are dichotomic questions, we can describe the opinion state of agents
as vectors in a two-dimensional Hilbert space. Both the eigenvectors of $\widehat{A}$ and $\widehat{B}$, defined by
equation (\ref{eig1}), form two orthonormal bases of the Hilbert space $H$. Thus we can express the eigenvectors of
$\widehat{A}$ in the basis of the eigenvectors of $\widehat{B}$, obtaining the general equations:
\begin{eqnarray}\label{A-B-basis}
|a_0\rangle=cos(\theta)|b_0\rangle+sin(\theta)e^{i\phi}|b_1\rangle\\\nonumber
|a_1\rangle=-sin(\theta)e^{-i\phi}|b_0\rangle+cos(\theta)|b_1\rangle\,\,,
\end{eqnarray}
and vice-versa
\begin{eqnarray} \label{B-A-basis}
|b_0\rangle=cos(\theta)|a_0\rangle-sin(\theta)e^{i\phi}|a_1\rangle\\\nonumber
|b_1\rangle=sin(\theta)e^{-i\phi}|a_0\rangle+cos(\theta)|a_1\rangle\,\,.
\end{eqnarray}
The transformations above are a change of basis, which can be described in terms of a unitary operator $\widehat{U}$
(element of $SU(2)$ group) such that $\sum_{ij}U_{ij}|a_i\rangle=|b_j\rangle$.
Moreover, they are useful to compute the conditional probabilities $P(a_1|b_1)=P(a_0|b_0)=cos^2(\theta)$ and
$P(a_1|b_0)=P(a_0|b_1)=sin^2(\theta)$.
It is important to note that, as evidenced in \cite{Rfranco_conj1}, in quantum mechanics we can not consider
simultaneously the two events $a_1$ and $b_1$; what we can consider are the conditional probabilities $P(a_1|b_1)$,
remembering that $P(a_1|b_1)P(a_1)$ could be different from $P(b_1|a_1)P(b_1)$. In other words, the elements of the two
basis of $A$ and $B$ should be handled carefully.

First of all, we consider a mixed state, that is an incoherent mixture of states $|a_i\rangle$ with probabilities
$|\alpha_i|^2$:
\begin{equation}\label{Srho-A}
\widehat{\rho}=|\alpha_0|^2|a_0\rangle \langle a_0| + |\alpha_1|^2|a_1\rangle \langle a_1|\,\,.
\end{equation}
We show that for this state the conjunction fallacy is not allowed: if we compute the probability $P(b_1)=\langle
b_1|\widehat{\rho}|b_1\rangle$, one obtains by using equation (\ref{A-B-basis}) the classical formula
\begin{equation}\label{no_interference}
P(b_1)=P(a_0)P(b_1|a_0)+P(a_1)P(b_1|a_1)\,\,.
\end{equation}
This equation is consistent with formula  (\ref{Conj_fallacy_classic}), and evidences that $P(b_1)$ can not be lower
than $P(a_1)P(b_1|a_1)$. Mixed states thus exhibit a behavior similar to the classic situation, without conjunction
fallacy.

Let us now consider as the initial state the following superposition
\begin{equation}\label{S-A}
|s\rangle=\alpha_0|a_0\rangle + \alpha_1|a_1\rangle
\end{equation}
where $\alpha_i$ are in general complex parameters such that $P(a_i)=|\alpha_i|^2$, reproducing the same statistical
predictions for $\widehat{A}$ of (\ref{Srho-A}). By using equation (\ref{A-B-basis}), we can express this state in the
basis of $\widehat{B}$, obtaining
\begin{equation}\label{S-b}
|s\rangle=[\alpha_0 cos(\theta)-\alpha_1 sin(\theta)e^{-i\phi}]|b_0\rangle + [\alpha_0 sin(\theta)e^{i\phi}+\alpha_1
cos(\theta)]|b_1\rangle.
\end{equation}
We now consider the probabilities $P(a_1)$, $P(b_1)$ and the conditional probability $P(a_1|b_1)$: from equation
(\ref{S-b}), we have that $P(b_1)$ is $|\alpha_0 sin(\theta)e^{i\phi}+\alpha_1 cos(\theta)|^2$, obtaining
\begin{equation}\label{interference}
P(b_1)=P(a_0)P(b_1|a_0)+P(a_1)P(b_1|a_1)+Re[\alpha_0 \alpha_1^*sin(2\theta)e^{i\phi}]
\end{equation}
The presence of the last term, known as the interference term
$I(s,A)$ can produce conjunction fallacy effects: in fact, if we
impose that $P(a_0)P(b_1|a_0)+I(s,A)<0$, we have
$P(b_1)<P(a_1)P(b_1|a_1)$. Thus the sign of the interference term
can determine the conjunction fallacy, while the parameter $\phi$
can give to this effect more or less strength. A positive
interference term enhances the prevalence of $P(b_1)$ on
$P(a_1,b_1)$, which can be considered a \textit{reverse
conjunction fallacy}.
The conjunction fallacy  can appear also for $P(a_1)$, if we write
the same initial state in the basis of $\widehat{B}$
\begin{equation}\label{S-BB}
|s\rangle=\beta_0|b_0\rangle + \beta_1|b_1\rangle
\end{equation}
The probability $P(a_1)$ can be written, with similar calculations, as
\begin{equation}\label{interference_b}
P(a_1)=P(b_0)P(a_1|b_0)+P(b_1)P(a_1|b_1)-Re[\beta_0 \beta_1^*sin(2\theta)e^{i\phi}]
\end{equation}
evidencing once again an interference term $I(s,B)$. If we want
the presence of conjunction fallacy $P(a_1)<P(b_1)P(a_1|b_1)$, we
 impose $P(b_0)P(a_1|b_0)+I(s,B)<0$.

We consider now the results presented in \cite{Myamoto}, where
several probability combination models for conjunction errors are
presented: we want to show that the use of quantum formalism
allows us to explain the experimental data (and in particular the
results of table III of \cite{Yates}) in a more complete way.
We consider for simplicity real superposition coefficients
$\alpha_i,\beta_j$, and $\phi=0$: interference effects can occur
also without complex numbers. Moreover, since
$|\alpha_0|^2+|\alpha_1|^2=1$, we can write $\alpha_0=a
cos(\theta_a), \alpha_1=a sin(\theta_a)$, with $a$ positive
number. Thus the basis transformation (\ref{A-B-basis}) leads to
the simple relations $\beta_0=a cos(\theta_a - \theta)$ and
$\beta_1=a sin(\theta_a - \theta)$. It is evident that the angle
$\theta$ controls the correlations between the questions $A$ and
$B$: for $\theta\simeq 0$ the answers $a_1$ and $b_1$ are strictly
correlated, for $\theta\simeq \pm\pi/4$ they are uncorrelated,
while for $\theta\simeq \pm\pi/2$ they are anti-correlated.
Similarly, $\theta_a$ controls the probabilities $P(a_i)$: if
$\theta_a\simeq 0$, then $P(a_1)\simeq 1$, and if $\theta_a\simeq
\pm \pi/2$, then $P(a_1)\simeq 0$. The presence of conjunction
fallacy for $P(b_1)$ and $P(a_1)$ entails, with such assumptions,
respectively
\begin{eqnarray}\label{conj_ab}
1+2tan(\theta_a)cotan(\theta)<0\\\nonumber
1-2tan(\theta_a-\theta)cotan(\theta)<0\,
\end{eqnarray}
where $cotan(x)=tan(x)^{-1}$.
These two formulas can take into account from a qualitative point of view the experimental results of \cite{Yates}: for
correlated questions ( $\theta\simeq 0$) the two inequalities are simultaneously satisfied for a range of $\theta_a$
such that $\theta_a\simeq \pi/2$ and $\theta_a-\theta>\pi/2$, which means a configuration of probability
$P(a_1)/P(b_1)$ high/high. This configuration also allows  a range of $\theta_a$ zero conjunction errors, for $\theta$
positive.
For anti-correlated questions ( $\theta\simeq \pm \pi/2$) the two
inequalities can not simultaneously satisfied; only one inequality
can be satisfied when $|\theta_a|\simeq  \pm\pi/2$, which means a
configuration of probability high/low or low/high.
Finally, for uncorrelated questions ( $\theta\simeq \pm \pi/4$) the two inequalities are simultaneously satisfied when
$\theta_a\simeq \pm \pi/2$, which means a configuration of probability high/high.

In \cite{Myamoto} other possible models to explain these data are presented; for example, the probability combination
models, where level of ratings of probability $R(A),R(B)$ are connected to the belief strength $S(A),S(B)$) through a
function $M$ nonlinear. A modified version of this model introduces the additional term $s_0$, which can be interpreted
as the initial impression. In \cite{Yates} has been purposed a signed summation of belief strength, reproducing some of
the the conjunction effects, but allowing for presence of self contradictory conjunction. Moreover, in \cite{Myamoto}
some arguments against the representativeness interpretation of conjunction fallacy are presented: the unrelated case,
in fact, seems to be unexplained by representativeness arguments.

Finally, we note that the experiments evidencing the conjunction fallacy show the rates of agents which have considered
the conjunction of the two events more probable than the single events: in other words, the experiments show how many
agents have considered $P(a_1,b_1)$ higher than $P(b_1)$ for example. Equation (\ref{interference}) entail that if all
the agents are in the same state (\ref{S-A}), then the rate of agents which exhibit conjunction fallacy is 100\%. To
solve this difference from experimental data, we note that we have used the hypothesis that all the agents are in the
same state (\ref{S-A}). However, the quantum formalism allows us to prepare the opinion state in a more general way:
for example, we can prepare agents in the state (\ref{S-A}) with a probability $P_1$, and in a state (\ref{Srho-A})
which does not exhibit conjunction fallacy with a probability $1-P_1$, thus reproducing  the experimental predictions.
At the moment, the experimental data are not enough to determine completely the initial state of the system: in fact,
we should measure not only the frequency of agents for which $P(a_1)$ is lower than $P(a_1,b_1)$, but also $P(a_1),
P(b_1)$ and the conditional probabilities $P(a_i,b_j)$.
\subsection{Other quantum approaches}
We note that a recent paper \cite{aerts3} contains a different attempt to describe a similar fallacy in terms of
quantum formalism. In particular, the experimental results of Hampton \cite{Hampton} are considered: given two concepts
$A$ and $B$ and an item $X$, the membership weights relevant to $A$ and $B$ ($\mu(A)$ and $\mu(B)$) are compared with
$\mu(A\, or\, B)$. The experiment of \cite{Hampton} evidences that in many cases $\mu(A \,or\, B)<\mu(A)$ and $\mu(A or
B)<\mu(B)$. This effect is called \textit{underextension} of the two concepts $A$ and $B$. The attempt of \cite{aerts3}
is to deduce this effect from the description of concept membership in terms of quantum formalism. In particular, the
situation of complete membership respect to the concept $A$ is described by the ket $|A\rangle$, while the opposite
situation (complete non-membership) by the orthogonal vector $|A'\rangle$. Similarly for the concept $B$, we have
$|B\rangle$ and $|B'\rangle$. The membership  to $A$ of an item $X$, which corresponds to the vector $|X\rangle$, is
given in this formalism by $|\langle A|X\rangle|^2=\mu(A)$. The ket $|X\rangle$ can be written in the basis $A-A'$ or
$B-B'$, obtaining
\begin{eqnarray}\label{Aerts-SAB}
|X\rangle=a e^{i\alpha}|A\rangle + a' e^{i\alpha'}|A'\rangle\\
|X\rangle=b e^{i\beta}|B\rangle + b' e^{i\beta'}|B'\rangle
\end{eqnarray}
We show now that the approach of \cite{aerts3}, even if good in its starting points, contains a mistake in the use of
the quantum formalism. The crucial point is the introduction of the vector $|AB\rangle=|A\rangle + |B\rangle$, which is
used to compute the membership of $|X\rangle$ relevant to the concept $A \, or \, B$: we expect that this vector
describes a composite concept which contains with equal weights the concept $A$ and $B$.  By using equation
(\ref{B-A-basis}), we can write $|B\rangle=sin{\theta}e^{-i\phi}|A'\rangle + cos(\theta)|A\rangle$. and thus $|A\rangle
+ |B\rangle$ should be interpreted as a superposition $\psi_a |A\rangle + \psi_a' |A'\rangle$, where
$\psi_a=1+cos(\theta)$: this is incorrect, since $|\langle AB|A\rangle|^2$ should be $1/2$. In other words,
\cite{aerts3} has obtained an underextension of concepts not only for the item $X$ but also for the constituent concept
$A$.

A second critique is that we could also consider $|A\rangle + e^{i\lambda}|B\rangle$, obtaining completely different
effects and leading to an ambiguity in the formalism; moreover, $|A\rangle$ and $|B\rangle$ are contained in two
different basis and it is important in the quantum formalism  to write each vector in terms of vectors of a single
basis.

We think that the right approach to the problem is to express $\mu(A \, or \, B)$ in terms of
$\mu(A)+\mu(B)-\mu(A\,and\,B)$. Even if the quantum formalism evidences problems in considering simultaneously states
of different basis, we can use equations (\ref{interference}) and (\ref{interference_b}) to show that $\mu(A)<\mu(A,B)$
or $\mu(B)<\mu(A,B)$, thus obtaining that $\mu(A\, or \, B)<\mu(A)$ or $\mu(A\, or \, B)<\mu(B)$ (connected to the
underextension effect). 
\section{New fallacies}\label{section:new}
The rich mathematical formalism of quantum mechanics and the interference effects allow us to predict or explain other
fallacies in the bounded-rationality regime. For example, 1) the ordering effects, 2) the disjunction effect, 3) the
conditional probability fallacy, 4) the framing effect and 5) the uncertainty effect. We give in this final section a
brief description of these effects, which will be described in other articles.

1) In the bounded-rationality regime, the order with which we consider two questions $A$ and $B$ is important (see also
the case of rational ignorance \cite{Rfranco_rat_ign_1}). Similarly to the repeated Stern-Gerlach experiment, we ask a
question relevant to the operator $\widehat{A}$, then a second question relevant to the non-commuting operator
$\widehat{B}$ and finally again $\widehat{A}$. Equation (\ref{joint_cond2}) defines the probability that the result of
the second question is $b_j$, given $a_i$. But the third measure leads to a non-null probability $P(a_k \rightarrow
b_j)$ for $k\neq i$: in other words the third question can give a result for the observable $\widehat{A}$ different
from the first.  The bounded rationality situation has manifested an irrational behavior of the agent. The question
$\widehat{A}$ has been asked two times, but what has been changed is the \textit{context}. The opinion state of the
agent in the test has evidenced a contextuality effect. Thus it has great importance, in bounded rationality, also the
temporal order of the different questions.

2) The disjunction effect is an intriguing phenomenon discovered by Tversky and  Shafir \cite{Shafir} with important
consequences in modelling the interactions between inference and decision. This effect, like the conjunction fallacy,
considers the probabilities relevant to two events which can be associated to two non-commuting operators. A first
attempt to give an explanation of the effect within the quantum formalism has been given by \cite{Busemeyer}, by
considering the two questions relevant to two different Hilbert spaces. The consequences of this approach are that we
obtain an entangled state, but also that the evolution of the initial state can lead to a state which contradicts the
initial information given to the agent. A new explanation of the disjunction effect will be given in a separate paper:
here we only observe that the conjunction fallacy can be applied to show how the perceived probability $P(b_1)$ (for
example) can be lower than $P(a_0)P(b_1|a_0)+P(a_1)P(b_1|a_1)$, because the interference terms like in equation
(\ref{interference}) may appear.

3) The framing effects may have a similar explanation of the conjunction and disjunction effect: the interference terms
in fact are able to lower the probability $P(b_j)$ when the initial information lead to an opportune pure state in a
basis relevant to a non-commuting operator.

4) The conditional probability fallacy is the assumption that $P(a_j|b_i)=P(b_i|a_j)$. In classical probability theory
this is not valid in general, while in quantum mechanics it is always true, as can be seen from equation
(\ref{conditional_p}).

5) Finally we make the hypothesis of the existence of an uncertainty effect, which is a consequence of the well-known
uncertainty principle in quantum mechanics. The experimental data of \cite{osherson} show that a not-null percentage of
agents in the tests fails in the implication $X -and- Y \models X,Y$. We argue that this percentage may change,
depending on the commutator of the associated operators $\widehat{X},\widehat{Y}$. The uncertainty principle in Hilbert
spaces with discrete dimension has been formulated in a more general form \cite{lur}: given the uncertainty of an
observable $\widehat{X}$, defined as the statistical variance of the randomly fluctuating measurement outcomes $
\delta^2(\widehat{A})=\langle \widehat{A}^2 \rangle - \langle \widehat{A} \rangle ^2$, the local uncertainty relations
\cite{lur} state that, for any set $\{\widehat{X},\widehat{Y},\widehat{Z},...\}$ of non-commuting operators, there
exist a non-trivial limit $U$ such that $\delta^2(\widehat{X})+\delta^2(\widehat{Y})+\delta^2(\widehat{Z})+... \geq U$.
This new form of the uncertainty principle may apply to bounded-rationality regime, leading to predictions similar to
those cited of \cite{osherson}: even if we try to prepare the opinion state of agents such that the uncertainty of $X$
and $Y$ is null, the sum $\delta^2(\widehat{X})+\delta^2(\widehat{Y})$ can not be null.

All these paradoxical effects in general are due to the usual belief that we can assign pre-defined elements of reality
to individual observables also in regime of bounded rationality. In a classical situation, if we ask to an agent the
two questions associated to the observables $\widehat{A}$, $\widehat{B}$, we can consider simultaneously the two
answers $(a_i,b_j)$, and we can study the joint probability $P(a_i,b_j)$. In a bounded rationality regime, instead,
this is not possible if the related observables are non-commuting. We say that the answers to these questions can not
be known contemporarily, thus giving an important limit to the complete knowledge of the opinion state of an agent in
bounded-rationality regime.

This effect in microscopic world is called quantum contextuality \cite{contextuality}, and evidences, for any
measurement,  the influence of other non-commuting observables previously considered.
%

%
%
\section{Conclusions}\label{conclusions}
This article, addressed both to quantum physicists and to experts of cognitive science, evidences the incompatibility
of quantum formalism with Bayes' rule of classic probability theory, by deriving the violation of equation
(\ref{Conj_fallacy_classic})  in bounded-rationality regime. In particular, we use mathematical objects like vector
state and density matrix to describe the opinion state of agents, and hermitian operators for the questions: in section
\ref{section:conjunction} we show that the conjunction fallacy can be explained as an interference effect when two
different questions (relevant to two non-commuting operators) are considered.

This seems to confirm the comment of \cite{osherson}, for which the conjunction fallacy seems to involve failure to
coordinate the logical structure of events with first impressions about chance. The first impressions about chance may
be encoded in the quantum phase, which leads to interference effects. In fact, we have seen that states (\ref{S-A}) and
(\ref{Srho-A}) do not differ for the statistical predictions of $A$, but for the presence of a phase, which gives us
the information of how the same superposition of states is considered initially by the agent.

Thus we conclude that the conjunction fallacy can be considered as a natural consequence of the quantum formalism used
to describe the bounded-rationality regime. By the way, the formalism introduced does not only give an explanation of
the fallacy, but also has a predictive character: in fact, we have predicted a reverse conjunction fallacy, for which
the probability of the conjunction of two events is much less than the probability assigned to a single event.
Moreover, in section \ref{section:new} we propose other effects which are consequences of the formalism introduced.

As evidenced in the conjunction fallacy, the experimental test evidencing these effects are
very difficult to be performed, because they must avoid collateral effects and the preparation of the opinion states
must be considered carefully.
\\\\\\
I wish to thank Jerome Busemeyer and Peter Bruza for making very useful comments and suggestions.
 \footnotesize
%
%

\end{document}